\documentstyle[preprint,aps]{revtex}
\begin{document}
\draft
\title{Multi-particle Correlations in Quaternionic Quantum Systems}
\author{S. P. Brumby\cite{brumby} %
and G. C. Joshi\cite{joshi}}
\address{School of Physics, University of Melbourne, %
Research Centre for High Energy Physics,\\
Parkville, Victoria 3052, Australia.}
\author{R. Anderson\cite{anderson}}
\address{Dibner Institute for the History of Science and Technology,\\
38 Memorial Drive, Massachusetts Institute of Technology, %
Cambridge, MA 02139, USA.}
\maketitle
\begin{center}
UM--P--94/54 ; RCHEP--94/15

hep-th/9406040
\end{center}
\begin{abstract}
We investigate the outcomes of measurements on correlated, few-body
quantum systems described by a quaternionic quantum mechanics that allows for
regions of quaternionic curvature.
We find that a multi-particle interferometry experiment using a correlated
system of four nonrelativistic, spin-half particles has the potential to
detect the presence of quaternionic curvature.
Two-body systems, however, are shown to give predictions identical to those of
standard quantum mechanics when relative angles are used in the construction
of the operators corresponding to measurements of particle spin components.
\end{abstract}
\narrowtext
\section{Introduction}
Quaternionic quantum mechanics (QQM) was raised as a possibility by Birkhoff
and von Neumann~\cite{BvN} in 1936, and has been elaborated since then by a
variety of authors using different theoretical approaches~\cite{FJSS,Per,Ad1,%
HzB,NaJ,Ad2,DMK,HzR}.
To date, however, the absence of clear experimental evidence has meant that
researchers have constructed models with a concern to ensure that the
quaternionic aspects are hidden in situations where standard, complex quantum
mechanics (CQM) is successful.
Fully quaternionic interactions are permitted, but the higher quaternionic
components of the wavefunctions in these models exponentially decay in the
absence of quaternionic-dissipative potentials~\cite{Ad2}.
Further, it is customary to work in what Finkelstein {\em et al.}~\cite{FJSS}
have classified as the Q--flat limit of the theory, thereby neglecting much of
the geometrical content of the full theory.

Our contention is that within the general theory of QQM (General QQM)~\cite%
{FJSS}, which is the simplest example of a truly general implementation
of QQM along the lines of Birkhoff's and von Neumann's work, the quaternionic
nature of the states can manifest itself in collective, nonlocal effects.

In particular, there emerges a possibility of testing the prediction of General
QQM that the $i$ of CQM becomes a field of pure imaginary unit quaternions
on spacetime, by the simultaneous measurement at remote points of the intrinsic
spin of particles in entangled states.
In order to relate the complex algebras at each point, Ref.~\cite{FJSS} uses
concepts from differential geometry; gauge connections, covariant derivatives,
and curvature have their quaternionic analogues in Q--connections, Q--covariant
derivatives, and Q--curvature.

To treat many-body systems in their nonmutually interacting state (i.e., to
construct Fock spaces of particles), we use the tensor product of quaternionic
Hilbert modules developed by Horwitz and Razon~\cite{HzR}.
This permits the definition of a scalar product of quaternionic multi-particle
wavefunctions, by abandoning linearity in each factor of the tensor product in
favour of a quotient-group structure, (we present a summary in the appendix).

In the next section, we examine the predictions of General QQM for the
the archetypical experimental test (Bohm~\cite{Bohm}) of the
Einstein--Podolsky--Rosen programme (EPR)~\cite{EPR}.
This was also the situation Bell~\cite{Bell} used in his famous demonstration
that CQM could not be completed in accord with the EPR paper's assumptions.

This case raises a number of foundational questions about how we encode
physical information in the specification of mathematical operators
corresponding to particular measurements.

We then consider a similar experiment on a correlated four-electron state
proposed by Greenberger {\em et al.}~\cite{GHSZ}.
It is in this situation that we expect manifestations of quaternionic
behaviour.

Given the enormous importance of the results engendered by Bell's work for
exploring the foundations of CQM, we view the results of our paper as a
preliminary contribution to locating their significance for  exploring the
foundations of QQM.

\section{Two Body Correlations}
Bohm's {\em ged\"{a}nkenexperiment\/} consists of simultaneous spin component
measurements on a system of two spin-half particles prepared in an entangled
state
\begin{equation}
|\Psi \rangle = \case{1}{\sqrt{2}}\bigl(|+\rangle _{1}|-\rangle _{2} -\ %
|-\rangle _{1}|+\rangle _{2}\bigr) .
\label{eq:2body}
\end{equation}

The preparation procedure, which is visualized as the decay out of a two-body
S--state, results in a state which is rotationally invariant.
That is, the complete entangled state is invariant under the group of rotations
$\{g\otimes g\,|\,g \in {\rm SU(2)}\}$, where ${\rm SU(2)}$ is the covering
group of ${\rm O}_3 $.
This is equivalent to stating that the measurement of internal spins of both
particles along the same spatial axis, is independent of the orientation of
that axis.
(Note that this use of the quaternions, related to the spin-half nature of
fundamental fermions, is located wholly within the domain of CQM).
It is this invariance which essentially means we cannot ascribe any physical
reality to definite spin components prior to their measurement.

The entangled character of the state is also invariant under this group,
which implies that under linear, continuous evolution in free space,
the state remains entangled and does not collapse into a trivial product of
wavefunctions at distinct points.

In addition, this invariance means that when we measure spin components along
different spatial directions, it is only the relative angular displacements
of the analysing directions that are physically significant.

In CQM, this state can be multiplied by an additional
arbitrary phase-factor which does not change the physical system (i.e.\ we
actually deal with a equivalence class of states, of which the above is merely
representative, the class relation being multiplication by a phase-factor).
This enables us to choose the initial state Eq.\ (\ref{eq:2body})
to be real, and in particular, to choose each factor to be real.

We now proceed to QQM, which is equivalent to a complete and distinct CQM at
each point of spacetime, related to each other through the Q--connection.
Hence it contains within it the property of local phase-invariance of states.
In this case, the choice of a real initial state corresponds to nulling the
phase-factors associated with each single particle wavefunction separately.

The ${\rm O}_3 $ invariance of the quaternionic system, in the sense of spin
measurements along the same spatial axis being independent of the orientation
of that axis, is assured by the local equivalence between CQM and QQM.
Hence the rotational invariance of the state still holds within QQM with the
theory therefore still implying that individual particle spins fail to have any
physical reality prior to their experimental resolution.

The Hermitian operator corresponding to the measurement, at position
${\bf x}$, of internal spin of a single particle, along a spatial unit vector
\begin{equation}
{\bf n} = (\sin \theta \cos \phi , \sin \theta \sin \phi , \cos \theta) ,
\end{equation}
where the angles are defined with respect to some laboratory frame of
reference, is

\begin{equation}
{\bf n} \cdot \bbox{\sigma} = \left( \begin{array}{cc}
\cos \theta & \sin \theta e^{-\phi \eta} \\
\sin \theta e^{\phi \eta} & -\cos \theta
\end{array}  \right) .
\end{equation}

Here, $\eta $ is the quaternionic generalisation of the imaginary unit of CQM
which, expanded with respect to a fixed basis of imaginary units, is
\begin{equation}
\eta = \eta({\bf x}) = \sum_{r=1}^3 h_r ({\bf x})i_r ,
\end{equation}
where $\forall {\bf x},\ {\bf h}({\bf x}) \in {\bf  R}^3,\ %
\|{\bf h}\| = 1$.  That is, QQM requires the outcomes of measurements to
depend on where they are carried out (phenomenon of Q--curvature).

The expectation value of the product of a component of each particles' spin,
\begin{equation}
E_{\Psi }({\bf n}, {\bf n}') \equiv \case{1}{2} ( P_{++} + P_{--} - P_{+-}%
- P_{-+}),
\end{equation}
which is equivalent to calculating
\begin{equation}
E_{\Psi }({\bf n}, {\bf n}') = \langle \Psi |\ {\bf n}\cdot %
\bbox{\sigma}_{(1)} \otimes {\bf n}'\cdot \bbox{\sigma}_{(2)}|\Psi \rangle ,
\end{equation}
where
\begin{equation}
P_{+-} = \langle + |\langle - |\ {\bf n} \cdot \bbox{\sigma}_{(1)}%
\otimes {\bf n}'\cdot \bbox{\sigma}_{(2)} |-\rangle |+\rangle .
\end{equation}
Note that the subscripts of $P$ refer to the first terms of each tensor
product of states.

With the detectors at ${\bf x}_1 $ and ${\bf x}_2 $ respectively, without %
loss of generality we set $\eta({\bf x}_1 ) = i_1 $ and drop the position
label from $\eta({\bf x}_2 )$.
\begin{eqnarray}
P_{+-}	& = & 	\sum_{(\rho )\in {\cal I}^{2}} [ \langle + | \langle - |\ %
      	(\sin \theta e^{\phi i_{1}}|+\rangle )_{\rho_{1}}(\sin \theta '%
      	e^{-\phi '\eta}|-\rangle )_{\rho_{2}}] \nonumber\\
      	&&\ \ \ \times (1\otimes 1\otimes 1\cdot 1_{D({\bf  H})}, %
	i_{\rho_{1}}\otimes i_{\rho_{2}}\otimes 1\cdot 1_{D({\bf  H})})%
	_{{\bf  H}} .\nonumber\\
\end{eqnarray}

For the reasons given previously, we consider the basis spin states
$|\pm \rangle $ to be real.
Then we can identify them with their formally real $0^{\text{th}}$ components
(in the sense of \cite{HzR})
\begin{equation}
(|\pm \rangle )_{\sigma} = \case{1}{4}\sum_{\varpi \in {\cal I}} i_{\varpi} %
(|\pm \rangle i^{\ast}_{\sigma}) i^{\ast}_{\varpi} = |\pm \rangle \delta%
_{\sigma 0}.
\end{equation}

Hence
\begin{eqnarray}
P_{+-} = & \sin\theta \sin\theta' & \sum_{\rho \in {\cal I}}%
	\:(e^{-\phi '\eta})_{\rho} \nonumber\\
	&&\ \times [ \cos\phi \,(1_{D({\bf  H})},\:1 \otimes i_{\rho}%
	\otimes 1\cdot 1_{D({\bf  H})}\bigr)_{\bf H} \nonumber\\
	&&\ \ + \sin\phi \,(1_{D({\bf  H})},\:i_1 \otimes i_{\rho}%
	\otimes 1\cdot 1_{D({\bf  H})}\bigr)_{\bf H} ]. \nonumber \\
\end{eqnarray}

Applying the recursive definition to the inner product of tensor producted
quaternionic algebras (given in the appendix), we have
\begin{equation}
(1_{D({\bf  H})}, 1 \otimes i_{\rho} \otimes 1\cdot 1_{D({\bf H})}\bigr)%
_{\bf  H} = \case{1}{3}i_{\rho} + \case{2}{3}\delta_{\rho 0},
\end{equation}
\begin{equation}
(1_{D({\bf  H})}, i_1 \otimes i_{\rho} \otimes 1\cdot 1_{D({\bf H})}\bigr)%
_{\bf  H} = \case{1}{3}\delta_{\rho 0}i_1 - \case{1}{3}\delta_{\rho 1}.
\end{equation}

Hence
\begin{eqnarray}
P_{+-} = \sin\theta \sin\theta' &[& \cos\phi \cos\phi' + \case{1}{3}%
	\sin\phi \sin\phi' h_{1} \nonumber \\
&&+ \case{1}{3}\sin\phi \cos\phi' i_1 - \case{1}{3}\cos\phi \sin\phi' %
	\eta ]. \nonumber
\end{eqnarray}

Similarly,
\begin{eqnarray}
P_{-+}  &=&    	\langle - |\langle + | {\bf n} \cdot \bbox{\sigma}_{(1)}%
	\otimes {\bf n}'\cdot \bbox{\sigma}_{(2)} |+\rangle |-\rangle ,%
	\nonumber \\
	&=&	\sin\theta \sin\theta'(\langle - |\otimes\langle + |,\:%
	e^{-\phi i_{1}}|+\rangle \otimes e^{\phi '\eta}|-\rangle )%
	_{\bf  H} \nonumber\\
	&=&	P_{+-}(\phi \rightarrow -\phi\:;\:\phi' \rightarrow -\phi'),%
	\nonumber
\end{eqnarray}
so that in combination, we obtain the real result
\begin{equation}
P_{+-} + P_{-+}  = 2\sin\theta \sin\theta' (\cos\phi \cos\phi' + \case{1}{3}%
\sin\phi \sin\phi' h_{1}).
\end{equation}

Now
\begin{equation}
P_{++} = \langle + |\langle - | {\bf n} \cdot \bbox{\sigma}_{(1)}%
\otimes {\bf n}'\cdot \bbox{\sigma}_{(2)} |+\rangle |-\rangle ,
\end{equation}
\begin{equation}
P_{--} = P_{++} = -\cos\theta \cos\theta' .
\end{equation}

Therefore we have the (real) expectation value
\begin{eqnarray}
E_{\Psi }({\bf n}, {\bf n}') &=& -\cos\theta \cos\theta' \nonumber \\
&&-\sin\theta \sin\theta'(\cos\phi \cos\phi' %
+ \case{1}{3}h_{1}\sin\phi \sin\phi'). \nonumber \\
\end{eqnarray}
The final term in this sum represents the manifestation of QQM in this
nonrelativistic, correlated quantum system.  It is a specialisation of the
Euclidean inner product
\mbox{${\bf h}({\bf x}_1 )\cdot {\bf h}({\bf x}_2 )$},
arising from the natural inner product structure imposed upon the quaternionic
algebra.

As stated before, rotational invariance of the initial entangled state
Eq.\ (\ref{eq:2body}) requires the
use of relative angles for the phases $\phi$, $\phi'$, (i.e., %
$\phi \rightarrow \phi_{\text{rel}} = \phi - \phi'$, $\phi' = 0$),
Hence, we retrieve the CQM result
\begin{equation}
E_{\Psi }^0 ({\bf n}, {\bf n}') = -\cos\theta \cos\theta' %
-\sin\theta \sin\theta'\cos\phi_{\text{rel}}.
\end{equation}
That is, by defining a polar axis, $\phi = \phi' = 0$, we have
\begin{equation}
E_{\Psi }({\bf n}, {\bf n}') = -\cos\theta_{\text{rel}}%
= -{\bf n}\cdot {\bf n}'.
\end{equation}
Alternatively, restricting the ${\bf n}$'s to a plane,
$\theta = \theta' = \pi/2 $, implies
\begin{equation}
E_{\Psi }({\bf n}, {\bf n}') = -\cos\phi_{\text{rel}}%
= -{\bf n}\cdot {\bf n}'.
\end{equation}

The fact that general QQM hides itself at this level of two-body systems
is an interesting result, especially as it tends to imply that QQM might not
manifest itself in models neglecting $\geq 3$-body interactions.  This would
suggest that the present lack of experimental evidence for QQM may be due to
the subtlety of the theory.

\section{Four Body Correlations}\label{sec:4body}

We now consider a 4-particle correlated system described by Greenberger
{\em et al.}~\cite{GHSZ}.
\begin{equation}
|\Psi _{\text{GHSZ}}\rangle =\case{1}{\sqrt{2}}\bigl(%
|+\rangle _{1}|+\rangle _{2}|-\rangle _{3}|-\rangle _{4} %
-|-\rangle _{1}|-\rangle _{2}|+\rangle _{3}|+\rangle _{4}\bigr).
\label{eq:GHSZwavefn}
\end{equation}

We carry out simultaneous (or at least, space-like separated) measurements of
a component of spin of each particle, and consider the product of these values.
The quantum mechanical expectation value is thus
\begin{equation}
E_{\text{GHSZ}}(\{{\bf n}_{(j)}\}) = \langle \Psi _{\text{GHSZ}}|%
\bigotimes_{j=1}^{4} {\bf n}_{(j)}\cdot \bbox{\sigma}_{(j)}%
|\Psi _{\text{GHSZ}}\rangle ,
\end{equation}

\begin{equation}
P_{+-} = \langle +|\langle +|\langle -|\langle -|%
\bigotimes_{j=1}^{4} {\bf n}_{(j)}\cdot \bbox{\sigma}_{(j)}%
|-\rangle |-\rangle |+\rangle |+\rangle ,
\end{equation}
\widetext
\begin{equation}
P_{+-} = \Bigl(\prod_{j=1}^{4}\sin\theta_j \Bigr) \sum_{(\rho ) %
\in {\cal I}^{4}} (e^{\phi_1 \eta({\bf x}_1 )})^{\rho_1 }(e^{\phi_2 %
\eta({\bf x}_2 )})^{\rho_2 }(e^{-\phi_3 \eta({\bf x}_3 )})^{\rho_3 }%
(e^{-\phi_4 \eta({\bf x}_4 )})^{\rho_4 }%
(1_{D({\bf  H})},\:{\sf i}_{(\rho)}\otimes 1 \cdot 1_{D({\bf H})})_{\bf H},
\end{equation}
where $\{\theta_j , \phi_j \}$ are the polar and azimuthal angles specifying
the direction ${\bf n}_{j}$ along which a measurement is made on the
$j^{\text{th}}$ particle.

We now simplify this expression by considering the special configuration
$\forall j$, $\theta_j = 0$ and without loss of generality take $\eta_3 = i_1$.
Further, we use relative angles:
\mbox{$\forall j$, $\phi_j \rightarrow \phi_j - \phi_4 $}.

Hence,
\begin{eqnarray}
P_{+-} = \sum_{(\rho ) \in {\cal I}^{2}} (e^{\phi_1 \eta({\bf x}_1 )})^{%
\rho_1 }(e^{\phi_2 \eta({\bf x}_2 )})^{\rho_2 }%
& \bigl[ & \cos\phi_3 \,(1_{D({\bf  H})},\:{\sf i}_{(\rho)}\otimes 1 %
\otimes 1\otimes 1\cdot 1_{D({\bf  H})})_{\bf H} \nonumber \\
&& - \sin\phi_3 \,(1_{D({\bf  H})},\:%
{\sf i}_{(\rho)}\otimes i_1\otimes 1\otimes 1\cdot 1_{D({\bf  H})})_{\bf H}%
 \bigr] .
\label{eq:4bdP}
\end{eqnarray}

Using the recursive definition for the inner-product of quaternion algebras,
we obtain from Eqs.~(\ref{eq:GHSZwavefn})--(\ref{eq:4bdP}),
\begin{eqnarray}
E_{\text{GHSZ}}(\{{\bf n}_{(j)}\}) & = & %
-\case{1}{2}(P_{+-} + P_{-+}) \nonumber \\
&=& -\cos\phi_1 \cos\phi_2 \cos\phi_3 %
- \case{1}{3}h_1 ({\bf x}_1)\sin\phi_1 \cos\phi_2 \sin\phi_3 %
- \case{1}{3}h_1 ({\bf x}_2)\cos\phi_1 \sin\phi_2 \sin\phi_3 \nonumber \\
&& + \case{1}{3}{\bf h}({\bf x}_1)\cdot{\bf h}({\bf x}_2)\sin\phi_1 %
\sin\phi_2 \cos\phi_3 .
\label{eq:4bdE}
\end{eqnarray}
\narrowtext

Hence, we find that QQM effects are manifest in experiments on this 4-body
correlated system.  Moreover, we expect QQM to give different predictions to
CQM for all $N \geq 2$ body system.  (On this point, we mention that Horwitz
and Razon~\cite{HzR} have suggested that as $N \rightarrow \infty$,
with their definition of a multi-particle inner-product, we recover the
quantized-field commutator relations of complex quantum field theory).

The CQM expectation value is
\begin{equation}
E_{\text{GHSZ}}^0 (\{{\bf n}_{(j)}\}) =%
- \cos ( \phi_1 + \phi_2 - \phi_3 )
\end{equation}

Comparing this with Eq.\ (\ref{eq:4bdE}), we see that the CQM limit is reached
by nulling the Q--curvature and removing the numerical factors that arise from
the recursive definition of the scalar product.
\begin{mathletters}
\begin{equation}
\forall {\bf x},{\bf x}'\!\!,\ \ \ %
\case{1}{3}{\bf h}({\bf x})\cdot {\bf h}({\bf x}') \rightarrow 1,
\end{equation}
\begin{equation}
\text{or,}\ \forall {\bf x},\ %
h_1 ({\bf x}) = \sqrt{3},\ h_2 ({\bf x})= h_3 ({\bf x})= 0,
\end{equation}
\end{mathletters}
where in the latter we have gone back to Eq.\ (\ref{eq:4bdP}) and taken
$h_1 ({\bf x}_3 ) = 1 \rightarrow \sqrt{3}$.  This ``rescaling'' of the
$\eta$ field might possibly be avoided by using a different definition for the
inner product, but the occurrence of ${\bf h}({\bf x})\cdot%
{\bf h}({\bf x}')$ coefficients to the relative-orientation dependent terms
in the expectation value might still be expected (the same terms arise in the
simplest approaches to this calculation~\cite{SPBHon}).

Note that if instead we considered an experiment which made use of plane
polarised photons, in the manner described by Ref.~\cite{GHSZ},
then because photons possess a single quantum of angular momentum, the
operator corresponding to a polarisation filter has real components~\cite{CSh}
and QQM has no opportunity to manifest itself.
This situation is changed by the use of circularly polarised photons, in which
case the quantum mechanical prescription for calculating the expectation
values necessarily uses complex numbers, giving the quaternionic entangled
state a probability distribution sensitive to a changing $\eta$--field
(hence, not predicted by CQM).

\section{Conclusion}
We see that it is an artifice of particle number which hides QQM in two-body
situations.  For more complicated systems, General QQM gives expectation
values that differ formally from those of CQM.
The importance of particle number in this situation is analogous to that
pertaining to correlations in CQM.
As Greenberger {\em et al.} have shown, Bell type inequalities
which permit local realistic hidden variable theories to agree with quantum
mechanical predictions in certain regions of parameter space (though not in all
regions) only occur for two-particle entangled states.  For more complicated
systems, however, CQM predicts correlations for entangled systems which are
unable to be described by local hidden variable theories.

Our result in Eq.\ (\ref{eq:4bdE}) indicates that experimental tests of
4-particle correlations have either the potential to reveal quaternionic
components or to set limits on their values.
Our prediction of quaternionic terms in multiparticle correlation experiments
provides a further motivation to the reasons given by Greenberger {\em et al.}
to undertake such experiments.

If such an experimental test of QQM were carried out, and no quaternionic
correction terms were discovered, then in some clear sense we would have
evidence against what we would maintain is the most natural interpretation
and implementation of the suggestion of Birkhoff and von Neumann.

QQM may still be the ``real'' theory of quantum mechanics, but its contribution
to phenomena at human scales, and in our vicinity of spacetime, may be slight.
In this case until there occurs a situation where QQM provides an explanation
analogous in significance to General Relativity's explanation of the anomalous
precession of Mercury, the theory will remain tentative.
(This raises an obvious question about what could act as a strong source of
Q--curvature, to which no answer has yet been proposed in the literature.)

A more modest perspective would be to sidestep for the moment questions as to
whether or not QQM is the ``real theory'' of quantum phenomena, and simply to
view QQM in a phenomenological sense as a theory whose extra parameters affords
a way of better capturing experiment results that have defied explanation by
other means.   The results of the type we present in Sec.~\ref{sec:4body}
indicate the manner in which this parametrization can occur.
Naturally any success on this score invites a deeper consideration on the
viability of QQM.

Depending on the outcomes of these possibilities, QQM may tell us something
crucial about the necessity of both q-numbers and c-numbers in the correct
description of physical phenomena as we know them.
QQM exists as a potential alternative theory because of the correspondence we
set up between experimentally observable quantities and Hermitian operators
(which necessarily possess real eigenvalues).
If experiments fail to support General QQM, then we have found out something
about the physical standing of non--Hermitian operators and unobservable
properties of a system.

We await experimental clarification of these concerns.

\acknowledgements
One of the authors, S.\ P.\ B., acknowledges the support of an Australian
Postgraduate Research Award.  G.\ C.\ J.\ was supported by the Australian
Research Council and the University of Melbourne.
R.\ A.\  acknowledges the support provided by the Dibner Institute, and is
grateful for the hospitality provided by the Institute.

\appendix
\section*{Tensor product of quaternion algebras}
\begin{eqnarray*}
\bf  R &\ \ \ \ &\text{the real algebra}\\
\bf  H &\ \ \ \ &\text{the quaternion algebra}\\
1,\,i_1 ,\,i_2 ,\,i_3 &\ \ \ \ &\text{a basis for $\bf  H$, where}%
\ \forall r, s \in \{1,\,2,\,3\}, \\
{} &\ \ \ \ & i_r i_s = -\delta_{rs} + \sum{}_{t=1}^{3}\epsilon_{rst}i_t \\
\rm Im\,\bf  H &\ \ \ \ &\text{the set $\{\eta\}$ of imaginary elements of %
$\bf  H$},\\
{} &\ \ \ \ &\eta = \sum{}_{r=1}^{3}h_r i_r \ \text{where}\ {\bf h} \in %
{\bf  R}^3,\ \|{\bf h}\|=1\\
{\cal H}_{\bf  H} &\ \ \ \ &\text{a Hilbert $\bf H$-module}\\
{\bf  C}_{\eta} &\ \ \ \ &\text{the complex algebra generated by 1 and}\ \eta\\
D &\ \ \ \ &\text{the real linear span}\ %
{\bf  H} \otimes \cdots \otimes {\bf H}\\
\cal I &\ \ \ \ &\text{the set}\ \{0,\,1,\,2,\,3\}\\
(\rho) &\ \ \ \ &\text{an element}\ (\rho_1 ,\ldots ,\,\rho_N) \in %
{\cal I}^{N}\\
{\sf i}_{(\rho)} &\ \ \ \ &\text{the tensor product}\ i_{\rho_1}%
\otimes \cdots \otimes i_{\rho_N}, \\
{} &\ \ \ \ &\text{where}\ i_0 = 1\\
{}[q] &\ \ \ \ &\text{the product}\ q \otimes \cdots \otimes q \in D\\
{}[q]_{j} &\ \ \ \ &\text{the product}\ 1\otimes \cdots \otimes q%
\otimes \cdots \otimes 1 \in D,\ \\
{} &\ \ \ \ &\text{with the $q$ in the $j\/$th place}\\
(q)_0 &\ \ \ \ &\text{the real component of}\ q \in {\bf  H}\\
(q)_\eta &\ \ \ \ &\text{the ${\bf  C}_{\eta}$ component of $q$}\\
({\sf q})_0 &\ \ \ \ &(q_1 )_0 \ldots (q_N )_0 \ %
\text{for}\ {\sf q} = q_1 \otimes \cdots \otimes q_N \in D\\
({\sf q})_\eta &\ \ \ \ &(q_1 )_\eta \ldots (q_N )_\eta \\
A_\eta &\ \ \ \ &\text{a left ideal of $D$ generated by the set}\\
&\ \ \ \ &\ \{ [\eta]_1 - [\eta]_j \ |\ 2 \leq j \leq N\}\\
D(\eta) &\ \ \ \ &\text{the quotient}\ D/A_\eta \\
1_D &\ \ \ \ & 1 \otimes \cdots \otimes 1 \in D \\
1_{D(\eta)} &\ \ \ \ & 1_D + A_\eta \\
A_{\bf  H} &\ \ \ \ & \bigcap{}_{\eta \in \rm Im\,\bf H}\:A_\eta \\
D({\bf  H}) &\ \ \ \ &\text{the quotient}\ D/A_{\bf H} \\
1_{D({\bf  H})} &\ \ \ \ & 1_D + A_{\bf H} \\
\end{eqnarray*}

We define pairwise multiplication in the product algebra,
\begin{eqnarray}
{\sf q}{\sf r} &=& (q_1 \otimes \cdots \otimes q_N )(r_1 \otimes %
\cdots \otimes r_N )\nonumber \\
&=& (q_1 r_1 \otimes \cdots \otimes q_N r_N ).
\end{eqnarray}

$({\sf q})_{\cal Q}$ is defined inductively for ${\sf q} \in D$ by
\begin{equation}
(q_1 )_{\cal Q} = (q_1 )_0 = {\rm Re}\,q_1,
\end{equation}
and for $N > 1$,
\begin{eqnarray}
({\sf q})_{\cal Q} = &\frac{1}{N + 1}& [(q_1 q_N \otimes \cdots \otimes %
q_{N-1})_{\cal Q} \nonumber\\
&&  + \cdots + (q_1 \otimes \cdots \otimes q_{N-1} q_N)_{\cal Q} \nonumber\\
&& + 2(q_N )_0 (q_1 \otimes \cdots \otimes q_{N-1})_{\cal Q} ].
\end{eqnarray}

\begin{equation}
({\sf q}_1 \cdot 1_{D({\bf  H})},\:{\sf q}_2 \cdot 1_{D({\bf H})})%
_{\bf  H} = \sum_{j=0}^{3}({\sf q}_1^{\ast}[i_{j}^{\ast}]_N {\sf q}_2)%
_{\cal Q} \:i_j .
\end{equation}

For each $\psi \in {\cal H}_{\bf  H}$, the formally real components of $\psi$
are defined at each point ${\bf x}$  by
\begin{eqnarray}
(\psi )_{\bf  H} &=& \sum_{\sigma \in {\cal I}} %
\bigl(\psi \otimes i^{\ast}_{\sigma}({\bf x})\bigr)_{\cal Q} %
i_{\sigma}({\bf x}) \\
&=&\sum_{\sigma \in {\cal I}}\psi_{\sigma }^{{\bf x}}i_{\sigma}({\bf x}).
\end{eqnarray}
We now choose to decompose each tetrad $i_{\sigma}({\bf x})$ as a rotation
from a standard basis $i_{\rho}$ (the rotation being an element of the group
${\rm I}_1 \oplus {\rm O}_3 $).
\begin{equation}
(\psi )_{\bf  H} = \sum_{\rho \in {\cal I}}%
\Bigl(\sum_{\sigma \in {\cal I}}\psi _{\sigma}^{{\bf x}}{\cal O}%
_{\sigma \rho}({\bf x})\Bigr)i_{\rho} %
= \sum_{\rho \in {\cal I}} \psi _{\rho} i_{\rho}.
\end{equation}

We can now form the quaternion tensor product of $N$ single particle systems,
which is itself a left module over the quaternionic algebra:
\begin{equation}
\Psi q = \psi^{1} \otimes \cdots \otimes \psi^{N} \otimes q \in %
{\cal H}_{\bf  H}^N \otimes {\bf H}.
\end{equation}
As usual, for an interacting system, the states of the $N$--body interacting
quantum system can be decomposed in the basis formed by the $N$--tensor
product of a basis ${\bf  B}({\cal H}_{\bf H})$ for ${\cal H}_{\bf H}$.

Using the decomposition into formally real components of each single particle
wavefunction,
\begin{eqnarray}
\Psi & = & \sum_{(\rho) \in {\cal I}^{N}} (\psi^{1}_{\rho_1}i_{\rho_1})%
\otimes \cdots \otimes (\psi^{N}_{\rho_N}i_{\rho_N}) \otimes 1 \cdot %
1_{D({\bf  H})} \nonumber\\
&=&\sum_{(\rho) \in {\cal I}^{N}} (\psi^{1}_{\rho_1}\otimes \cdots \otimes %
\psi^{N}_{\rho_N})\otimes({\sf i}_{(\rho)} \otimes 1\cdot 1_{D({\bf  H})}),%
\nonumber \\
\end{eqnarray}
where we have taken the formally real components and formed them into a
correspondingly ordered $N$--tuple of functions.

Thus, $\Psi \in [{\cal L}^{2}({\bf  R}^d )]^{N} \otimes D^{(N+1)}({\bf H})$,
where ${\cal L}^{2}({\bf  R}^d )$ is the set of real, Lebesgue
square-integrable functions on $d$-dimensional Euclidean space
(for particles in a box, these functions are of finite support).

We can now define a scalar product of $N$--body wavefunctions:
\begin{eqnarray}
\langle \Psi , \Phi \rangle_{\bf  H} =%
\sum_{(\rho),(\sigma) \in {\cal I}^{N}} &&%
\langle \psi^{1}_{\rho_1}\otimes \cdots \otimes %
\psi^{N}_{\rho_N} | \phi^{1}_{\sigma_1}\otimes \cdots \otimes \phi^{N}%
_{\sigma_N}\rangle \nonumber\\
&& \!\!\!\times ({\sf i}_{(\rho)} \otimes 1\cdot 1_{D({\bf  H})},\:%
{\sf i}_{(\sigma)} \otimes 1\cdot 1_{D({\bf  H})})_{\bf H}. \nonumber \\
\end{eqnarray}

\end{document}